# LATTICE PHENOMENOLOGY

C.T. Sachrajda

*Physics Department, The University*
*Southampton, SO9 5NH, United Kingdom*

## ABSTRACT

A review of the status of lattice simulations in particle physics phenomenology is presented. Recent computations of leptonic decay constants of light and heavy mesons, and of the Isgur-Wise function relevant for semi-leptonic decays of $B$-mesons, are discussed in some detail. Calculations of other quantities are briefly outlined. The systematic errors inherent in lattice simulations, and procedures to reduce and control them, are described.

## 1 Introduction

In this talk I will briefly review some of the important questions in particle physics phenomenology which are being studied using the lattice formulation of (Quantum Chromodynamics) QCD and numerical simulations. I hope to convince you that "Lattice Phenomenology" is already providing important contributions in particle physics, and that it is developing into the premier quantitative tool for non-perturbative field theory. For some physical quantities lattice computations have been performed successfully for several years, and the emphasis is now on controlling and reducing the systematic errors in these computations. These quantities include the leptonic decay constants of mesons discussed in sections 2 and 3 below. For other quantities, such as the evaluation of the Isgur-Wise function of semi-leptonic decays of heavy mesons discussed in section 4, lattice studies are only just beginning.

Before presenting some recent results, perhaps it is necessary to explain briefly what lattice calculations are, and to discuss the major sources of uncertainty in these computations. The starting point for lattice studies is the evaluation of the functional integral:

$$I(x_1, x_2, \cdots, x_n) = \frac{1}{Z} \int [dA_\mu][d\psi][d\bar\psi] e^{-S} O(x_1, x_2, \cdots, x_n) \quad (1)$$

where $A$ and $\psi$ represent gluon and quark field respectively, $O(x_1, x_2, \cdots, x_n)$ is a multilocal operator composed of the fields, $S$ represents the action and $Z$ is the partition function. $I(x_1, x_2, \cdots, x_n)$ is the vacuum expectation value of the operator $O$. The functional integral $I$ is evaluated by discretising space and time, and generating field configurations weighted by the Boltzman factor $e^{-S}$. The physical quantities which can be studied depend on the choice of operator $O$. For example, by choosing $O$ to be a bilocal operator of the form $O(x_1, x_2) = J_h(x_2) J_h^\dagger(x_1)$, where $J_h^\dagger$ and $J_h$ are interpolating operators which can create or annihilate the hadron $h$, the propagation of hadron $h$ is studied, allowing one to evaluate its mass and the matrix element $\langle 0 | J_h | h \rangle$. By evaluating 3-point correlation functions with $O(x_1, x_2, x_3) = J_{h_2}(x_3) X(x_2) J_{h_1}^\dagger(x_1)$, where $X$ is some local operator, we can evaluate matrix elements of the form $\langle h_2 | X(0) | h_1 \rangle$. For many fundamental quantities in particle physics, particularly in studies of hadronic structure and weak decay processes, the (non-perturbative) strong interaction effects can be expressed as operator matrix elements of this kind. Hence the importance of lattice simulations.

I will attempt to suppress any discussion of the technology of lattice computations, nevertheless it is difficult for me to avoid using two pieces of lattice jargon. The first is $\beta \equiv 6/g_0^2(a)$, where $g_0$ is the bare coupling constant and $a$ is the lattice spacing. It is convenient in lattice simulations to fix $\beta$ (and hence $g_0(a)$), and to determine the corresponding value of the lattice spacing by comparing the lattice prediction for some physical quantity to its physical value, (rather than the apparently more natural procedure of fixing the lattice spacing and determining the corresponding value of $g_0$). The second piece of jargon is $\kappa$, the Wilson hopping parameter, which is proportional to the coefficient of $\bar\psi\psi$ in Wilson's discretisation of the quark action (as well as generalisations of this like the $SW$ action mentioned below). $\kappa$ is therefore a measure of the quark mass, and the "critical" value of $\kappa$, which corresponds to zero renormalised quark mass (and zero pion mass), is called $\kappa_c$.

The numerical computation of the functional integral in eq.(1) leads to an evaluation of operator matrix elements "from first principles". However, there are a number of approximations in these calculations, leading to uncertainties in the final results. First of all we have the "statistical" errors, that is the errors due to the fact that we are estimating the functional integral by sampling the integrand at a finite number of field configurations. The size of these errors can be estimated, using standard statistical methods, by observing how the result varies as configurations are added or removed. As will be clear from the results presented below, for many quantities (on typical lattices) it appears that the statistical errors are adequately small for 50-100 field configurations (although there are some important exceptions to this0.

More problematical are the systematic errors. These include "finite volume effects", i.e. errors due to the fact that the integrals are evaluated with space-time taken to be finite. These can be studied by repeating the simulations on lattices of different sizes, and this is now being done more frequently. Theoretically, for a range of interesting quantities, it is known that finite volume effects decrease exponentially with the volume [1], and numerically the effects appear to be small on currently used lattices, at least for quenched calculations. However for this to be the case the numerical simulations are performed with light quark masses which are heavier than the physical ones (typically corresponding to pions with masses in the range of about 400 MeV - 1 GeV), and the results are then extrapolated to the physical limit (which is essentially the chiral limit for which the quarks are massless). I mention in passing that the dependence of masses and energy levels on the spatial volume of the lattice can be used to measure scattering lengths [1].

A second source of systematic error is due to the finiteness of the lattice spacing. Again these errors can be studies by performing high statistics runs on lattices with different values of the lattice spacing (i.e. at different values of $\beta$). An example of this will be presented in sec.2. Another approach is that of "improvement" [2] in which the discretisation errors are formally reduced through the use of an "improved" action and operators. Some of the results presented below have been obtained with the use of the fermion-action proposed by Sheikholeslami and Wohlert (the $SW$ or "clover" action) [3]. With this action, and with the use of improved operators, the discretisation errors are reduced from those of $O(a)$ (present in simulations with Wilson fermions) to ones of $O(\alpha_s a)$ [4]. An exciting possibility is that, by the use of the renormalisation group transformations, it may be possible to construct "perfect actions", for which the discretisation errors will effectively be eliminated, and yet which will be practicable for numerical simulations [5]. Whether this will be possible should become apparent during the next year or so.

Finally there is the problem of "quenching". The results presented below have been obtained by neglecting the effects of quark loops (but including all gluonic effects), and it is difficult in general to estimate the error this induces. Where the results of quenched calculations can be compared to experimental data, they are generally in good agreement suggesting that these errors are moderate (an important exception may be the hyperfine splittings in quarkonia). Sometimes it is possible to use the renormalisation group equations to relate two physical quantities, and the difference between results obtained by using zero flavours or the physical number of flavours in quark loops provides an estimate of the effects of quenching. However for the lattice computations to be treated as a truly quantitative technique, the effects of quark loops should be included, for which better algorithms or theoretical developments (such as the perfect actions mentioned in the preceeding paragraph) are needed. Otherwise we will have to wait for several years (5-10 ?) while the improvements in computing resources become sufficiently powerful to cope with the evaluations of the fermion determinants, necessary for full QCD computations.

In this short talk I clearly have to be selective in the subjects I cover. I hope nevertheless that the topics I have chosen will provide an accurate picture of the status of computations in lattice QCD, illustrating both what has been, and is being, achieved, and also some of the outstanding difficulties and attempts to overcome them. I will not have time to discuss many interesting lattice computations for field theories other than QCD. I also regret that I will not be able to compare the lattice results with those from model calculations or QCD sum rules.

## 2 Decay Constants of Light Mesons

The leptonic decay constants of light mesons, (e.g. $f_\pi$ and $f_\rho$) are obtained from matrix elements which are among the simplest ones to compute using lattice QCD. They are defined by

$$<0|A_\mu(0)|\pi(p)> = f_\pi p_\mu \quad (2)$$

and

$$<0|V_\mu(0)|\rho(p)> = \epsilon_\mu m_\rho^2 / f_\rho \quad (3)$$

where $\epsilon_\mu$ is the polarisation vector of the $\rho$-meson, and $A$ and $V$ are the axial-vector and vector currents respectively. The experimental values of the decay constants are $f_\pi = 132$ MeV ($f_\pi/m_\rho \simeq 0.17$), and $1/f_\rho = 0.28(1)$.

These decay constants have been computed for several years now. During the last year there have been a number of high statistics evaluations, and I will restrict my discussion to these studies. In particular the GF11 Collaboration [6] have evaluated the decay constants at several values of the lattice spacing with sufficient precision to be able to attempt an extrapolation to the continuum limit $a = 0$. In Fig.1 I present the lattice results for $f_\pi/m_\rho/Z_A$ obtained by the GF11 [6] and APE [7] collaborations, using Wilson fermions *. $Z_A$ is the renormalisation constant relating the lattice axial current to the physical one, and is calculable in perturbation theory. In addition to the directly measured values, I have also included in Fig.1, the results obtained by extrapolating these values to physical quark and pion masses. The results from the two collaborations are clearly in excellent agreement. The lattice sizes and numbers of gluon configurations used in each simulation are presented in table 1. The values of the inverse lattice spacings presented in table 1 were obtained from the measured values of the mass of the $\rho$-meson from the corresponding simulation.

---

*The results presented here for the GF11 collaboration are ones which have been corrected since their original preprint was circulated

Figure 1: Values of $f_\pi/m_\rho/Z_A$ measured by the GF11 and APE Collaborations.

| Group | $\beta$ | $a^{-1}$ GeV | Lattice Size | Configs |
|---|---|---|---|---|
| GF11 | 5.70 | 1.36(2) | $16^3 \times 32$ | 219 |
| GF11 | 5.93 | 2.00(5) | $24^3 \times 36$ | 217 |
| APE | 6.00 | 2.23(5) | $18^3 \times 32$ | 210 |
| GF11 | 6.17 | 2.78(5) | $30 \times 32^2 \times 32$ | 219 |

Table 1: Parameters of the Simulations by the GF11 and APE collaborations whose results for the decay constants are presented in this talk

The results in Fig.1 show a surprisingly mild dependence on the lattice spacing. In order to extract the physical quantity $f_\pi/m_\rho$, they need to be multiplied by the renormalisation constant $Z_A$ (which also depends mildly on the lattice spacing). Following the suggestion of Lepage and Mackenzie [8], I use the expression $Z_A \simeq (1 - 0.316/(4\pi\beta)(8\kappa_c)^4)/8\kappa_c$, which includes the effects of a partial summation of tadpole diagrams (which are lattice artefacts giving large coefficients in lattice perturbation theory) [9, 10, 11]. The $Z_A$'s ob-tained in this way are 0.67 at $\beta = 5.7$, 0.74 at $\beta = 5.93$, 0.75 at $\beta = 6.0$ and finally 0.77 at $\beta = 6.17$. The results for $f_\pi/m_\rho$ obtained using these values of $Z_A$ are presented in table 2. It must be stressed that the errors given in table 2 are statistical errors only. From these results we see that the dependence on the lattice spacing is remarkably small for this particular quantity, and that the results are in excellent agreement with the physical value.

| $\beta$ | $f_\pi/m_\rho$ | $1/f_\rho$ |
|---|---|---|
| 5.70 | 0.176(7) | 0.41(2) |
| 5.93 | 0.165(5) | 0.36(2) |
| 6.00 | 0.167(7) | — |
| 6.17 | 0.165(5) | 0.34(2) |

Table 2: Values of $f_\pi/m_\rho$ and $1/f_\rho$ obtained by the GF11 and APE collaborations.

The situation is a little different for the decay constant of the $\rho$-meson. Following the analogous procedure to that for $f_\pi/m_\rho$ above, we obtain (from the data of the GF11 collaboration) the values of $1/f_\rho$ in the third column of table 2. If the behaviour with the lattice spacing is linear, the results for $1/f_\rho$ extrapolate to about $1/f_\rho = 0.25(3)$ at $a = 0$ [6]. If this is the case, then clearly there are significant $O(a)$ effects in the evaluation of $1/f_\rho$ at around $\beta \simeq 6.0$-6.2 for Wilson fermions, however further work is needed to establish that the behaviour with $a$ is indeed linear all the way down to $\beta = 5.7$.

Similar studies are beginning also with the $SW$-action, for which the discretisation errors are formally reduced [7, 12]. For example the UKQCD collaboration [12] find $f_\pi/m_\rho = 0.14(1)$ at $\beta = 6.2$ and the APE collaboration find $f_\pi/m_\rho = 0.16(1)$ at $\beta = 6.0$. Over the coming months it will become possible to study the $a$-dependence in some detail. I should also mention that it is practicable, in simulations using the $SW$ fermion action, to determine the renormalisation constants $Z_A$ and $Z_V$ non-perturbatively, by imposing the chiral Ward Identities [13], removing one source of uncertainty.

## 3 Decay Constants of Heavy Mesons

In this section I will review the status of computations of the decay constants of heavy-light pseudoscalar mesons, i.e. mesons with a heavy quark (anti-quark) and a light anti-quark (quark). The decay constant $f_B$ (in conjunction with the $B$-parameter of $B^0$-$\bar{B}^0$ mixing) is an unknown parameter needed for the determination of the elements of the $CKM$ matrix, and the $CP$ violating phase in particular. I will present results obtained using two different approaches. The first method is an extension of the calculations described in section 2 but with the mass of the heavy quark in the region of that the charm quark (for quarks which are much heavier than this, the Compton wavelengths are smaller than the lattice spacing). I will refer to this method as simulations with propagating heavy quarks. The second method involves simulations of the Heavy Quark Effective Theory directly on the lattice, and I will refer to this method as simulations with static heavy quarks. I will not have time in this talk to describe lattice studies in heavy quark physics using the non-relativistic formulation of QCD (see however item c in section 5).

Simulations with propagating heavy quarks have been performed for several years now, and for example, at the 1989 conference on Lattice Field Theory [14], Steve Sharpe summarised the results for $f_D$ as

$$f_D = 180 \pm 25 \pm 30 \, \text{MeV} \qquad (4)$$

Results from recent simulations all lie in this range.

In the heavy quark effective theory the decay constant of a heavy pseudoscalar ($P$) meson is predicted to behave as a function of its mass as follows:

$$f_P = \frac{A}{\sqrt{M_P}} \left[ (\alpha_s(M_P))^{-2/\beta_0} [1 + O(\alpha_s)] + O(1/M_P) \right] \qquad (5)$$

In lattice simulations it is the constant $A$ which is evaluated, following the method proposed by Eichten [15]. I will denote by $f_B^{\text{stat}}$ the value of $f_B$ obtained in this way, i.e. obtained by dropping the $O(1/M_B)$ terms. Early results for $f_B^{\text{stat}}$ gave surprisingly large values:

$$f_B^{\text{stat}} = 310 \pm 25 \pm 50 \, \text{MeV} \quad \text{ref.[16]} \qquad (6)$$
$$f_B^{\text{stat}} = 366 \pm 22 \pm 55 \, \text{MeV} \quad \text{ref.[17]} \qquad (7)$$

Clearly these results could only be made consistent with those in eq.(4) if there were large negative $O(1/M_P)$ corrections in the charm region, and significant ones for the $B$-meson, so that $f_B \neq f_B^{\text{stat}}$. It was therefore important to check whether this was the case, and it was found in simulations with propagating quarks that indeed the quantity $f_P/\sqrt{M_P}(\alpha_s(M_P))^{2/\beta_0}$ does increase as $M_P$ is increased [18, 19].

A compilation of the results for $f_B^{\text{stat}}$ is presented in table 3 [20]. In the last column I present the results as they were quoted in the publications, but I also present the values of the lattice spacing and renormalisation constant which were used to obtain the result. Part of the reason for the spread of results is due to different procedures, particularly in the choice of $Z_A^{\text{stat}}$, but there is still some debate whether all of the collaborations can isolate the contribution of the ground state sufficiently accurately. To make further progress, more work on improving the evaluation of $Z_A^{\text{stat}}$ and above all on isolating the ground state is needed, and is in progress. There is some preliminary evidence that $f_B^{\text{stat}}$ may be decreasing as $a \to 0$ [23, 24], and it will be very interesting to be able to check this when the precision of the calculations improves.

This year two groups have presented new results obtained with propagating heavy quarks. Bernard, Labrenz and Soni, have performed simulations at $\beta = 6.3$ (20 configurations on a $24^3 \times 55$ lattice) and at $\beta = 6.0$ (19 configurations on a $16^3 \times 39$ lattice and 8 configurations on a $24^3 \times 39$ lattice) using Wilson fermions. These authors attempt to reduce the systematic errors associated with $O(m_Q a)$ effects (where $Q$ is the heavy quark) by modifying the heavy quark propagator following a procedure based on the structure of the free-field propagator [25, 26], (it will be very interesting to see whether this procedure for reducing $O(m_Q a)$ effects in simulations with Wilson fermions can be established when quantum loops are included, and this work is in progress [27]). Bernard, Labrenz and Soni quote

$$f_B = 187 \pm 10 \pm 34 \pm 15 \, \text{MeV} \qquad (8)$$
$$f_D = 208 \pm 9 \pm 35 \pm 12 \, \text{MeV} \qquad (9)$$

These values are based on their results with both propagating and static heavy quarks.

The second collaboration to present results this year with propagating quarks is the UKQCD collaboration which performed simulations at $\beta = 6.2$ (60 configurations on a $24^3 \times 48$ lattice), and $\beta = 6.0$ (36 configurations on a $16^3 \times 48$ lattice) using the $SW$ fermion action. Since the leading discretisation errors for the $SW$ action are of $O(\alpha_s m_Q a)$ (instead of $O(m_Q a)$ as for Wilson fermions), it was important to check that the results found with Wilson fermions [18, 19], and the dependence on $m_Q$ in particular, are reproduced [†]. In Fig.2 I show the results for the "scaling" quantity $f_P/\sqrt{M_P}(\alpha_s(M_P))^{2/\beta_0}$ as a function of $1/M_P$ obtained by the UKQCD collaboration from their simulation at $\beta = 6.2$. The open points correspond to the measured values at three values of the mass of the light quark (decreasing as the mass of the light quark is decreased) and the solid points are the results obtained after extrapolation to the chiral limit. The solid line is a linear fit to the heaviest three of the four points, and the broken line is a quadratic fit to all four points. The increase of this quantity as $M_P$ inreases is manifest. Also on this figure is shown the value of $f_B^{\text{stat}}$ obtained from 20 of the 60 configurations, which is about 2-3 standard deviations above the value obtained by extrapolation from the results with propagating quarks. From their results

---

[†]UKQCD have also analysed 18 (of the 36) configurations on the lattice at $\beta = 6.0$ using Wilson fermions, and compare the results for the two different fermion actions

| Ref. | Action | $\beta$ | $a^{-1}$ GeV | $Z_A^{stat}$ | $f_B^{\text{stat}}$ MeV |
|---|---|---|---|---|---|
| [21] | Wilson | 5.9 | 1.75 | 0.79 | $319 \pm 11$ |
| [16] | Wilson | 6.0 | $2.0 \pm 0.2$ | 0.8 | $310 \pm 25 \pm 50$ |
| [17] | Wilson | 6.0 | $2.2 \pm 0.2$ | 0.8 | $366 \pm 22 \pm 55$ |
| [7] | Wilson | 6.0 | $2.11 \pm .05 \pm .10$ | 0.8 | $350 \pm 40 \pm 30$ |
| [7] | SW | 6.0 | $2.05 \pm 0.06$ | 0.89 | $370 \pm 40$ |
| [20] | SW | 6.0 | $2.0 {}^{+3}_{-2}$ | 0.78 | $286 {}^{+8}_{-10} {}^{+67}_{-42}$ |
| [20] | SW | 6.2 | $2.7 {}^{+7}_{-1}$ | 0.79 | $253 {}^{+16}_{-15} {}^{+105}_{-14}$ |
| [22] | Wilson | 6.3 | $3.21 \pm .09 \pm .17$ | 0.69 | $235 \pm 20 \pm 21$ |
| [23] | Wilson | 5.74, 6.0, 6.26 | 1.12, 1.88, 2.78 | 0.71(8) | $230 \pm 22 \pm 26$ |

Table 3: Compilation of Lattice Results for $f_B^{\text{stat}}$

Figure 2: Values of $f_P/M_P^{1/2} (\alpha_s(M_P)/\alpha_s(M_B))^{(2/\beta_0)}$ measured by the UKQCD Collaboration.

with propagating quarks UKQCD quote:

$$f_B = 160 {}^{+6}_{-6} {}^{+53}_{-14} \text{ MeV} \qquad (10)$$

$$f_D = 185 {}^{+4}_{-3} {}^{+42}_{-7} \text{ MeV} \qquad (11)$$

The reason for such asymmetric errors in eqs.(10) and (11) is the uncertainty in the value of the lattice spacing obtained from the physics of light hadrons, for which the UKQCD collaboration quote $a^{-1} = 2.7 {}^{+7}_{-1}$ GeV.

My summary of the lattice results for the decay constants of heavy mesons, based on the above and earlier simulations, are

$$f_B = 180 \pm 40 \text{ MeV} \qquad (12)$$

$$f_D = 200 \pm 30 \text{ MeV} \qquad (13)$$

An important step now will be to study the dependence of the results obtained with the SW fermion action on the lattice spacing, and this work is in progress [28].

I would like to conclude this section by briefly mentioning a number of related quantities. First of all, as is clear from fig.2 the decay constants decrease as the mass of the light quark decreases (this is interpreted as being due to the fact that the size of the meson increases, so that the wavefunction at the origin is reduced). Lattice simulations typically give a result 10-20% larger for $f_{B_s}$ and $f_{D_s}$ than for $f_B$ and $f_D$ (the errors in the ratio are small), and for example UKQCD quote $f_{D_s} = 212 {}^{+4}_{-4} {}^{+46}_{-7}$ MeV, to be compared to the experimental results of $f_{D_s} = 232 \pm 45 \pm 20 \pm 48$ MeV [29] and $f_{D_s} = 344 \pm 37 \pm 53 \pm 42$ MeV [30].

Another important quantity is the (renormalisation group invariant) B-parameter for $B^0 - \bar{B}^0$ mixing for which the ELC collaboration find $f_B\sqrt{B_B} = 220 \pm 40$ MeV and $(f_{B_s}^2 B_{B_s})/(f_B^2 B_B) = 1.19 \pm 0.10$

## 4 The Isgur Wise Function

Semi-leptonic decays of heavy mesons are an important set of processes for studies of the standard model, and in the determination of the elements of the Cabibbo-Kobayashi-Maskawa (CKM) matrix (particularly for the $V_{cb}$ matix element). In the heavy quark effective theory, the two form-factors for the decay $B \to D$ + leptons and the four form-factors for decay $B \to D^*$ + leptons, are all given in terms of a single unknown function of $\omega$, ($\omega \equiv v_B \cdot v_D$, where $v_B$ and $v_D$ are the four-velocities of the $B$ and $D$ or $D^*$ mesons) [31]. Moreover this function, $\xi(\omega)$, known as the Isgur-Wise function [31], is normalised at the zero-recoil point, $\xi(1) = 1$. Experimental studies of these decays give results for $|V_{cb}|\xi(\omega)$ for values of $\omega > 1$, so a determination of $V_{cb}$ requires an extrapolation of the experimental results to $\omega = 1$. Lattice simulations can provide a determination of $\xi(\omega)$ and also test whether the charm quark is sufficiently massive for the heavy quark effective theory to be useful for $B \to D$ and $D^*$ decays.

A convenient way of determining $\xi(\omega)$ is by evaluat-

Figure 3: A comparison of the ARGUS experimental data for $|V_{cb}|\xi(\omega)$ with the lattice results from the UKQCD Collaboration.

ing the elastic matrix element

$$< D(p')|\bar{c}\gamma^\mu c|D(p) >= (p+p')^\mu f^+(q^2) \qquad (14)$$

where $q = p - p'$ and the single form-factor $f^+$, when considered as a function of $\omega$ ($q^2 = 2m_D^2(1-\omega)$), is equal to the Isgur-Wise function (once radiative corrections are included). The matrix element in eq.(14) can be evaluated using the techniques which have been developed for decays of charmed mesons (i.e. for $D \to K$ or $\pi$ and $D \to K^*$ or $\rho$ semi-leptonic decays) [32, 33]. Two groups have recently presented results for $\xi(\omega)$, Bernard, Shen and Soni [34] and the UKQCD collaboration [35, 36].

In fig.3 I present the results from the UKQCD collaboration, obtained at $\beta = 6.2$ from 60 configurations using the $SW$-action, and after extrapolation to the chiral limit for the light quark masses. Fitting the lattice results to Stech's relativistic-oscillator parametrisation [37]:

$$\xi(\omega) = \frac{2}{\omega+1} \exp\left(-(2\rho^2-1)\frac{\omega-1}{\omega+1}\right) \qquad (15)$$

the UKQCD collaboration find $\rho^2 = 1.2 \, {}^{+\,7}_{-\,2}$. Also on Fig.3 are the data points from the ARGUS experiment, and by comparing the lattice determination of $\xi(\omega)$ to the ARGUS data, the UKQCD collaboration find $V_{cb}\sqrt{\tau_B/1.48ps} = 0.038 \, {}^{+\,2}_{-\,2} \, {}^{+\,8}_{-\,3}$, where the first set of errors is due to the experimental uncertainty and the second is due to the uncertainty in the lattice determination of $\rho^2$. An extension of this work to another value of $\beta$ is in progress.

Bernard, Shen and Soni perform simulations with Wilson fermions, (but modifying the heavy quark propagator as discussed in section 3) at $\beta = 6.0$ (19 configurations on a $16^3 \times 39$ lattice and 8 configs on a $24^3 \times 39$ one) and $\beta = 6.3$ (20 configurations on a $24^3 \times 61$ lattice) They fix $\vec{p}'$ to be $\vec{0}$, and limit their calculations to the two lowest values of $|\vec{p}|$, however from their simulations at two values of $\beta$ and two values of the heavy quark mass they are able to obtain results for a range of $\omega$'s. These authors quote $\rho^2 = 1.41 \pm 0.19 \pm 0.19$ based on their results for the largest values of the light quark mass (an improvement in statistics is needed to control the extrapolation to the chiral limit). It would be very interesting to extend the range of momentum values in this work so as to be able to study better the systematic errors.

So far $\xi(\omega)$ has only been determined from the $D \to D$ matrix element in eq.(14). A very interesting set of checks will be performed during the next few months to establish whether the expectations of the heavy quark effective theory for the relation between the form factors in Pseudoscalar $\to$ Pseudoscalar and Pseudoscalar $\to$ Vector transitions are satisfied, for heavy quark masses in the region of the mass of the charm quark.

## 5 Some Other Lattice Studies in Particle Physics Phenomenology

I would like to end this talk by briefly mentioning a few other important studies in particle physics phenomenology which I haven't had time to discuss in detail.

a) Determination of the Strong Coupling Constant: El-Khadra et al. have calculated the strong coupling constant by computing the $1S$-$1P$ mass splitting in charmonium [38], finding

$$\alpha_{\overline{MS}}(5 \text{ GeV}) = 0.174 \pm 0.012 \qquad (16)$$

Lüscher et al., are attempting to compute $\alpha_s$ by using the size of the lattice as the dimensionful parameter[39].

b) Potentials: The potentials between static heavy quarks can be computed accurately [40, 41] and used to determine $\alpha_s$.

c) Quarkonia: One of the exciting recent developments has been the use of non-relativistic QCD for studies of the spectrum and properties of Quarkonia [42, 43]. Kronfeld and Mackenzie are also trying to generalise this approach to construct a field theory which would be applicable for all quark masses [25, 26], and it will be very interesting to see whether this can be achieved.

d) $B_K$: The parameter $B_K$, which parametrises the strong interaction effects in $K^0$-$\bar{K}^0$ mixing has been calculated by several groups during recent years. Lusignoli et al. [44] summarise the lattice results as $\hat{B}_K = 0.8 \pm 0.2$, where $\hat{B}_K$ is the renormalisation group invariant $B_K$-parameter. The most precise results came from Kilcup et al. [45], using the staggered formulation of lattice fermions, who performed simulations at several values of $\beta$ and extrapolated their results to zero lattice spacing. Recently (indeed since this talk was presented), Sharpe has demonstrated that the discretisation errors in these calculations are of $O(a^2)$ and not $O(a)$ [46], considerably reducing the uncertainty in the extrapolation to the continuum limit. The authors of ref.[45] quote $\hat{B}^K = 0.825 \pm 0.027 \pm 0.023$ (preliminary).

e) $\Delta I = 1/2$ Rule: There has been no progress on this very fundamental problem recently. The problem (for Wilson fermions) is how to subtract accurately the $1/a^3$ diveregence which occurs when the operators of dimension 6 which contain the strong interaction effects for this process, mix with the dimension 3 operator $\bar{s}d$ (through the so called eye-diagram). Although the subtraction can be performed in principle, the final results have errors of $O(100\%)$.

f) Semi-Leptonic Charm Decays: The semi-leptonic decays $D \to K, \pi, K^*$ or $\rho$ + leptons have been studied for several years now [32, 33, 47], giving pleasing results both for the form-factors at zero momentum transfer, and for the behaviour of the form-factors with momentum transfer. This year the ELC Collaboration has presented results at $\beta = 6.4$ with Wilson fermions [48], and although these have larger statistical errors than some of the earlier ones, the results are consistent (e.g. for the $D \to K$ decay ELC quote $f^+(0) = 0.65(18)$). In ref.[48] use of the heavy quark effective theory is made to develop a method for extrapolating the results to those for $B$-decays.

g) $b \to s\gamma$: At this conference we have heard from the CLEO collaboration about their observation of the process $B \to K^*\gamma$ [49]. In the standard model this process occurs through "penguin" diagrams, and the rate is sensitive to possible new physics. In order to determine whether the observed rate is consistent with the prediction of the standard model it is necessary to evaluate the strong interaction effects. This is being done by two lattice groups, Bernard, Hsieh and Soni [50] and the UKQCD collaboration [51], both using propagating heavy quarks. In fig.4 we see the measured results from the UKQCD collaboration for the form-factor $T_1$ which contains the strong interaction ef-fects for this process. The open squares are results obtained assuming that $T_1$ is independent of the mass of the light-quark, whereas the diamonds are the results obtained after extrapolation of the data to the chiral limit (there is no observed dependence on the light quark mass, but the errors grow significantly if an extrapolation is attempted). The crossed square and diamond are the corresponding points after linear extrapolation in the heavy quark mass to the physical value of $m_{K^*}/M_P$ ($M_P$ is the mass of the heavy pseudoscalar). The value deduced from the CLEO data, assuming that $m_t$ = 150 GeV (the dependence on $m_t$ is mild) is also shown, as is the extrapolated value from the simulation of Bernard, Hsieh and Soni (BHS). The lattice results suggest that the observed rate is consistent with the standard model contribution.

Figure 4: A comparison of the form-factor $T_1$ determined from the CLEO measurement and lattice computations.

## 6 Conclusions

I hope that I have managed in this talk to demonstrate that lattice simulations are making important contributions to particle physics phenomenology, and in particular to many processes which are under intensive experimental investigation. I have also tried to explain a little about the systematic errors present in the calculations and about attempts to reduce and control them. Lattice simulations are developing into the major quantitative tool for non-perturbative strong interaction physics.

## Acknowledgements

I am grateful to Don Weingarten for access to the data from the GF11 collaboration, and to Claude Bernard and Amarjit Soni for clarifying discussions of their results. It is a pleasure to thank my colleagues from the ELC and UKQCD groups for such fruitful and stimulating collaborations. I acknowledge the support of the UK Science and Engineering Research Council through the award of a Senior Fellowship. Finally I would like to thank the organisers of this conference for creating such a stimulating and enjoyable environment for scientific presentations and discussions.

## References

[1] M.Lüscher, Comm. Math. Phys. 104 (1986) 177; Comm. Math. Phys. 105 (1986) 153
[2] K.Symanzik, Nucl. Phys. B226 (1983) 187 and 205
[3] B.Sheikholeslami and R.Wohlert, Nucl. Phys. B259 (1985) 572
[4] G.Heatlie et al., Nucl. Phys. B352 (1991) 266
[5] P.Hasenfratz and F.Niedermeyer, University of Bern Preprint BU-93-17 (1993)
[6] GF11 Collaboration, F.Butler et al., IBM Yorktown Heights Preprint IBM-HET-93-3 (1993)
[7] APE Collaboration, C.R.Allton et al., University of Rome Preprint 928 (1993)
[8] G.P.Lepage and P.B.Mackenzie, Fermilab Preprint 91-355-T-Revised (1992)
[9] B.Meyer and C.Smith, Phys. Lett. B123 (1983) 62
[10] G.Martinelli and Y.C.Zhang, Phys. Lett. B123 (1983) 433
[11] R.Groot, J.Hoek and J.Smit, Nucl. Phys. B237 1984 111
[12] The UKQCD Collaboration, C.R.Allton et al., Edinburgh University Preprint 93-524 (1993)
[13] G.Martinelli, S.Petrarca, C.T.Sachrajda and A.Vladikas, Phys. Lett. B311 (1993) 241
[14] S.Sharpe, Nucl. Phys. B(Proc.Suppl.)17 (1990) 146
[15] E.Eichten, Nucl. Phys. B(Proc.Suppl.)4 (1988) 170
[16] C.R.Allton et al., Nucl. Phys. B349 (1991) 598
[17] C.Alexandrou et al., Phys. Lett. B256 (1991) 60
[18] C.R.Allton et al., Nucl. Phys. B(Proc.Suppl.)20 (1991) 504
[19] A.Abada et al., Nucl. Phys. B376 (1992) 172
[20] The UKQCD Collaboration, R.M.Baxter et al., Southampton University Preprint SHEP 92/93-24
[21] A.Duncan et al., Nucl. Phys. B(Proc.Suppl.)30 (1993) 433
[22] C.W.Bernard, J.Labrenz and A.Soni, University of Washington Preprint UW/PT-93-06 (1993)
[23] C.Alexandrou et al., PSI Preprint PSI-PR-92-27 (1992)
[24] E.Eichten, private communication
[25] A.S.Kronfeld, Nucl. Phys. B(Proc.Suppl.)30 (1993) 445
[26] P.B.Mackenzie, Nucl. Phys. B(Proc.Suppl.)30 (1993) 30
[27] A.S.Kronfeld and P.B.Mackenzie, private communication
[28] G.Martinelli, private communication
[29] The WA75 collaboration, S.Aoki et al., Prog. of Theoretical Physics 89 (1993) 131
[30] The CLEO collaboration, D.Acosta et al., Cornell Preprint CLNS-93-1238 (1993)
[31] N.Isgur and M.Wise, Phys. Lett. B232 (1989) 113; Phys. Lett. B237 (1990) 527
[32] C.W.Bernard, A.X.El-Khadra and A.Soni, Phys. Rev. D43 (1991) 2140; Phys. Rev. D45 (1992) 869
[33] M.Crisafulli et al., Phys. Lett. B223 (1989) 90; V.Lubicz, G.Martinelli and C.T.Sachrajda, Nucl. Phys. B356 (1991) 301; V.Lubicz, G.Martinelli, M.S.McCarthy and C.T.Sachrajda, Phys. Lett. B274 (1992) 415
[34] C.W.Bernard, Y.Shen and A.Soni, Boston University Preprint, BUHEP-93-15 (1993); Nucl. Phys. B(Proc.Suppl.)30 (1993) 473
[35] The UKQCD collaboration, S.P.Booth et al., Southampton University Preprint, SHEP 92/93-17 (1993)
[36] L.P.Lellouch, these proceedings
[37] M.Neubert and V.Rieckert, Nucl. Phys. B382 (1992) 97; M.Neubert, V.Rieckert, B.Stech and Q.P.Xu, in *Heavy Flavours*, Eds. A.J.Buras and M.Lindner (World Scientific, Singapore, 1992)
[38] A.X.El-Khadra et al., Phys. Rev. Lett.69 (1992) 729
[39] M.Lüscher et al., DESY Preprint 93-114 (1993); 92-157(1992); Nucl. Phys. B389 (1993) 247
[40] G.S.Bali and K.Schilling, Phys. Rev. D46 (1992) 2636; Phys. Rev. D47 (1993) 661
[41] The UKQCD Collaboration, S.P.Booth et al., Phys. Lett. B294 (1992) 385
[42] G.P.Lepage and B.A.Thacker, Phys. Rev. D43 (1991) 196
[43] C.Davies, J.Phys.G: Nucl. Part. Phys. 18 (1992) 1661
[44] M.Lusignoli, L.Maini, G.Martinelli and L.Reina, Nucl. Phys. B369 (1992) 139
[45] R.Gupta, G.Kilcup and S.Sharpe, Nucl. Phys. B(Proc.Suppl.)26 (1992) 197
[46] S.Sharpe, presented at the 1993 International Conference on Lattice Field Theory.
[47] T.Bhattacharya, D.Daniel and R.Gupta, Los Alamos Preprint, LA-UR-93-5580 (1993)
[48] A.Abada et al., University of Rome Preprint, 946-1993 (1993)
[49] R.Ammar et al., Phys. Rev. Lett.71 (1993) 674
[50] C.W.Bernard, P.F.Hsieh and A.Soni, Nucl. Phys. B(Proc.Suppl.)26 (1992) 347; Washington University Preprint HEP/93-35 (1993)
[51] UKQCD Collaboration, K.C.Bowler et al., University of Southampton Preprint SHEP93/94-01 (1993)